# A Heuristic Solution to Protect Communications in WDM Optical Networks using P-cycles

Hamza Drid [1], Bernard Cousin [1] and Miklos Molnar [2]
IRISA
[1] Université de Rennes I – Campus de Beaulieu 35042 Rennes Cedex, France
[2] INSA Département informatique F35043 Rennes Cedex, France
{hdrid, bcousin, molnar}@irisa.fr

### Abstract

**Optical WDM mesh networks are able to transport huge amount of information. The use of such technology however poses the problem of protection against failures such as fibre cuts. One of the principal methods for link protection used in optical WDM networks is pre-configured protection cycle (p-cycle). The major problem of this method of protection resides in finding the optimal set of p-cycles which protect the network for a given distribution of working capacity. Existing heuristics generate a large set of p-cycle candidates which are entirely independent of the network state, and from then the good sub-set of p-cycles which will protect the network is selected. In this paper, we propose a new algorithm of generation of p-cycles based on the incremental aggregation of the shortest cycles. Our generation of p-cycles depends on the state of the network. This enables us to choose an efficient set of p-cycles which will protect the network. The set of p-cycles that we generate is the final set which will protect the network, in other words our heuristic does not go through the additional step of p-cycle selection.**

## I. Introduction

Survivability becomes an important issue and well studied topic in wavelength division multiplexing (WDM) optical mesh networks. The WDM technology allows the transmission of huge amounts of data over the same fibre. Currently an optical fibre of 140 wavelengths is able of transmitting up to 14 Tbit/s over a distance of about 160 kilometres [11]. Consequently, any cut of such a fibre may lead to huge data loss and a lot of traffic being blocked. For this reason, methods and mechanisms of protection should be implemented to minimize the data loss when the failure occurs. Several methods for network protection have been proposed. One of the principal protection methods used in optical WDM networks is pre-configured protection cycles (p-cycle). The concept of p-cycle was introduced by Grover and Stamatelakis [2]. The basic idea of p-cycle protection has been inspired from the ring[1] protection. The main difference between p-cycle and ring protection is that, p-cycle protection not only protects the cycle links but protect all links whose end nodes belong to the p-cycle. In other words, p-cycle protects the cycle links and the straddling links in the cycle. A straddling link is a link which does not belong to the p-cycle but whose end-nodes are both on the p-cycle. Each p-cycle can offer two restoration paths to the failed straddling links without requiring any additional spare capacity. This propriety reduces effectively the required protection capacities. The major problem of this method of protection resides in finding the optimal set of p-cycles which protects the network for a given working capacity distribution. The p-cycle design can be formulated as a non-joint or joint optimization problem. In the first approach, after the working paths are routed (e.g. using shortest paths) the optimal set of p-cycles is calculated using available capacity [1][2][3][4]. In the second approach, the routing of the working paths (for the demands) and their p-cycles are computed simultaneously minimizing the total capacity [10][6][7].

Several solutions have been proposed in the literature to solve this optimization problem. These solutions can be classified into two categories: exact solution and heuristic solution. The first category uses generally the Integer Linear Programming (ILP) to find the optimal solution. However, ILP becomes unsuitable as the size of the network increases. Because the number of p-cycles possible in a graph grows exponentially as the size of the network grows (number of nodes and edges). The second category of solution is heuristic which is itself divided into two sub-classes: a heuristic approach based on an ILP formulation and a pure heuristic approach. In the first sub-class of solution, the limit set of p-cycle candidates is generated, and then the good sub-set of p-cycles which protects the network is selected using ILP formulation. The second sub-class (pure heuristic) tries to find a good solution without using ILP formulation. The objective of this solution is to reduce the time required to compute the good set of p-cycles that protects the network.

[1] In the ring protection, the traffic on the failed link is rerouted around the ring on the protection fibers between the nodes adjacent to the failure.

In this paper, we focus on the heuristic algorithms, which first generate a set of p-cycle candidates, and then a good sub-set of p-cycles that protects the network is selected. As we will describe in the next section, most of the solutions proposed in the literature consider that the p-cycles which have more straddling link as the most efficient p-cycles. Consequently, their generation of p-cycles is based only on the topology of the network and it is completely independent of the state of the network (working capacity distribution). An others drawbacks of these solutions are that they generate a big number of candidate p-cycles and the number of p-cycles selected for protection is very big. That makes configuration of network very complex. In this paper we propose a new heuristic for p-cycles generation that takes into account the state of network and that generates a small number of p-cycle which leads to good performance.

The rest of the paper is organized as follows. In section 2, we review some existing solutions and outline the limitations of each algorithm. In section 3, we describe our proposed algorithm of p-cycles generation based on the incremental aggregation. We evaluate our solution and we compare it with other algorithm in section 4. A conclusion is given in section 5.

## II. State of the Arts

In [8] Zhang and have proposed Straddling Link Algorithm (SLA). The main idea of this algorithm resides in the method of generation of the set of p-cycle candidates. The p-cycles generated by SLA contain potentially one straddling link. In order to construct this type of p-cycle, SLA looks for two shortest node-disjoint paths between the two end nodes of a link. If these paths exist, SLA combines these two paths to construct a p-cycle. SLA is very fast but the set of candidates cycles generated are generally inefficient because they have only one straddling link.

Doucette et .al [3] have proposed Capacitated Iterative Design Algorithm (CIDA) which is independent of ILP. This algorithm has two steps. In the first step, a set of candidate p-cycles is generated. In CIDA the generation of p-cycles is based on the transformation of the set of cycles provided by SLA into more efficient p-cycles. Three algorithms of transformations were proposed. The main principle of these algorithms consists of replacing the on-cycle[2] link of SLA p-cycle by a path between the end nodes of this link. After this transformation the on-cycle link becomes a straddling link of the new p-cycle obtained. In the second step, the efficient p-cycles are selected one by one iteratively from the set of candidate p-cycles. In each iteration step, the working capacity protected by a selected cycle is removed. The second step ends when all working capacities are protected. Efficient p-cycles are the cycles that have a high value of actual efficiency (AE).

The actual efficiency is defined as: $$AE(p) = \frac{\sum_{\forall i \in E} w_i \times X_{p,i}}{\sum_{i \in E / X_{p,i}=1} C_i}$$

Where $w_i$ is the amount of unprotected working capacity on link $i$, $X_{p,i}$ the number of protection paths provide by cycle $p$ after failure of link $j$. $C_j$ is the cost of link $j$.

Another algorithm has been proposed by [9], called Weighted DFS-based Cycle Search (WDCS). The goal of the algorithm is to compute a small set of candidate p-cycles that can lead to good performance when used by ILP or the heuristic algorithm. WDCS generates two types of cycle, high efficiency cycles and the shortest cycles. A high efficiency cycle is generated using DFS (Depth First Search). The main idea of WDCS is to construct a cycle which starts from a node x and finish at the same node by traversing the DFS path. WDCS assigns weights to the directed edges in the graph to control the behaviour of DFS and to find a p-cycle which includes more straddling links. To extend the searching path, DFS in this algorithm selects a neighbour node w of node v such that (v, w) has the highest weight among all outgoing links incident to v. In the basic DFS, when extending the search path from a node v, the neighbours of v are selected in arbitrary order. The number of cycles produced by WDCS is controlled by an input parameter k.

[2] On-cycle is a link that belong to the cycle

The second type of cycle is shortest cycle. This algorithm computes two shortest cycles for each link ***e*** in the graph. One of the cycles has ***e*** as a straddling link (generated with an algorithm similar to SLA) while the other has the link ***e*** as an on-cycle link. To create the second cycle, the algorithm finds the shortest path between the end-nodes of e in the graph and combines it with **e.** As in CIDA, the drawback of the WDCS algorithm is that the generation of candidate cycles is based only on the network topology without considering the state of the network (working capacity distribution).

As describe earlier, most of the solutions proposed in the literature consider the p-cycles which have more straddling link as the most efficient p-cycles. Consequently, their generation of p-cycles is based only on the topology of the network and it is completely independent of the state of the network (working capacity distribution). But in practice, however, the optimal solution can contain different sizes (number of straddling links) of p-cycle as shown in Fig. 1.

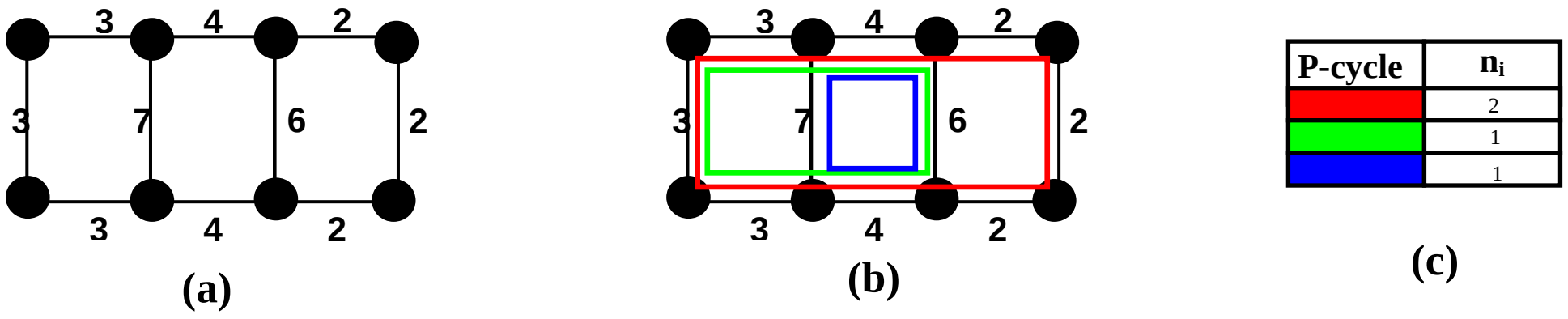

Fig. 1 : p-cycles example

Figure.1 (a) shows a WDM network with 8 optical switch, where each link has an integer value that indicates the working capacity of the link. Figure.1 (b) shows the set of optimal p-cycles that protect the network with a given working capacity. We see well that the optimal solution contains different kinds of p-cycles. This means that the p-cycles which have more number of straddling links are not always the best p-cycles. The table shown in Fig.5 (c) contains the final set of p-cycles which protect network where $n_i$ presents a capacity of p-cycle.

The solution that we propose in this paper generates p-cycles by taking into account the state of the network. This enables us to choose an efficiency set of p-cycles which will protect the network. Our solution is based on the incremental aggregation of the cycles. This consists of choosing the cycle with maximal protection capability among all shortest cycles in the network. Then we aggregate it iteratively with other shortest cycles (according to rules which take into account the state of the network) in order to obtain an efficient p-cycle. With our algorithm we can obtain p-cycles with a maximal number of straddling links. But for the reasons mentioned above our solution find p-cycles of different number of straddling links according to the state of network.

## III. Proposed Heuristic Approach

We model the WDM optical network as an undirected graph G =(V, E) where each node in V represents an optical switch and each edge in E represents a network link. Each link has $w_j$ working wavelength channels. A p-cycle *i* of a capacity $n_i$ can protect $n_i$ wavelengths on any on-cycle link of the p-cycle and $2\times n_i$ wavelengths on any straddling link of the p-cycle.

*Algorithm of P-cycle generation*

Our algorithm of generation of p-cycles has two steps. In the first step, the set of shortest cycles is generated. In the second step, we generate high efficiency p-cycles will be directly used to protect the network. Our generation of high efficiency p-cycles is based on incremental aggregation of shortest cycles.

*Step 1:*

The first step consists of generating all shortest cycles in the graph. A shortest cycle is an elementary cycle without a straddling link. The basic rule to generate such cycles is as follows: for each link X, we find the shortest path between the end nodes of X in G that is different than X.

Then we combine the path found with ***X*** to create a shortest cycle. When this step finish we delete all redundant shortest cycles.

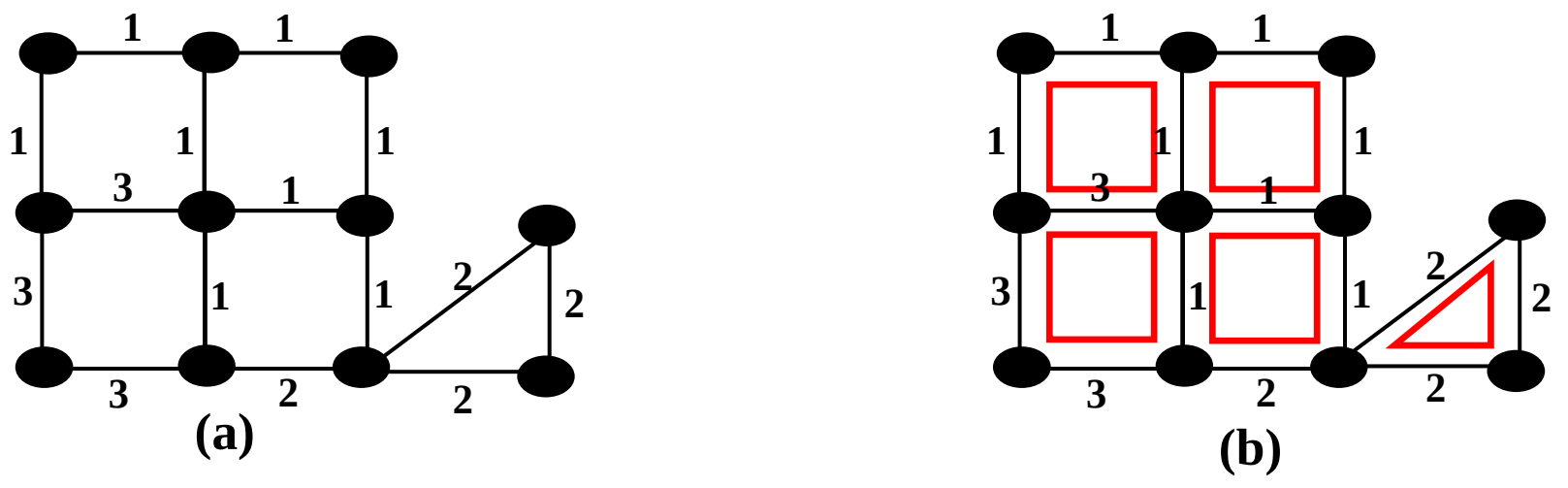

Fig. 3: Shortest cycles

After having found the set of shortest cycles, we try to simplify the graph with deleting the parts of graph which contain shortest cycles which cannot be improved. In other words, we delete the shortest cycles that will not have the straddling links. To perform this operation, we delete the shortest cycles that do not share any link with other cycles in the graph. The deleted shortest cycles will be put in the final set of p-cycles that protects network.

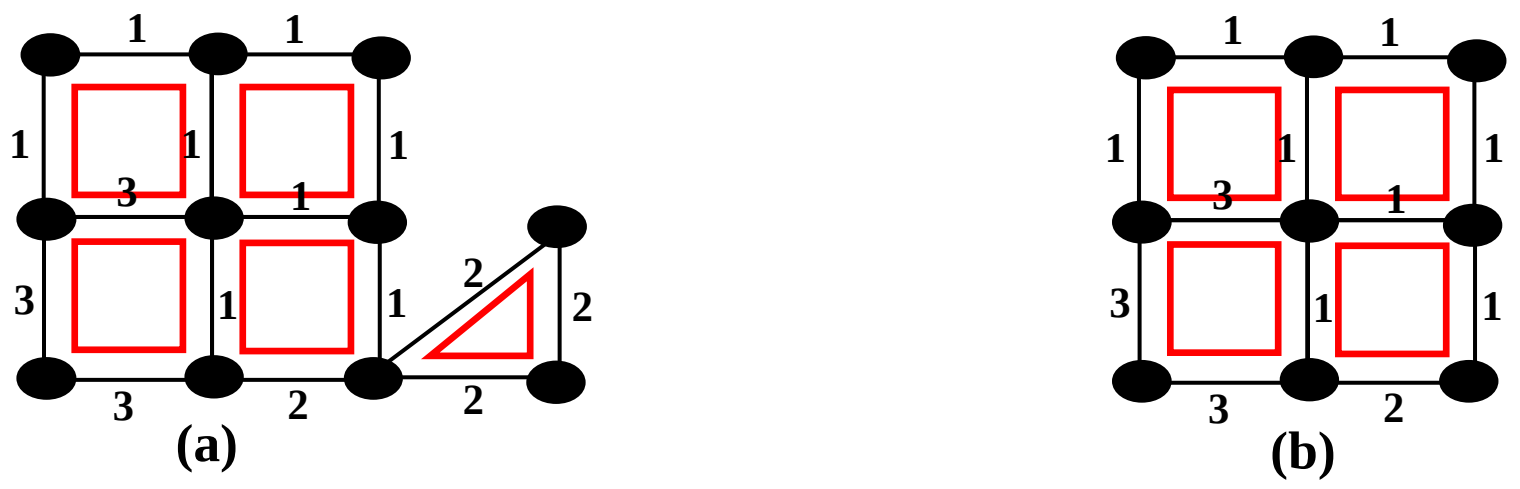

Fig. 4 Graph simplification

Figure.4 (b) shows the simplified graph obtained after removing the shortest cycles that are not improvable.

*Step 2:*

At the end of the first step, we obtain a simplified graph including only shortest cycles which could be improved. The second step starts in order to generate a set of high efficiency p-cycles. Our solution is based on the incremental aggregation of the shortest cycles, which consists in choosing a shortest cycle and by according rules described below we aggregate it with another cycle in order to obtain an efficient p-cycle.

At the beginning of our algorithm, we choose a cycle that has a link with a minimal working capacity $w_j$ among all shortest cycles in the network such that $w_j$ is greater than 0. We call the chosen cycle Cyc. Then, we try to aggregate Cyc with another shortest cycle $Cyc_i$. The aggregation is possible if $Cyc_i$ shares one and only one link with Cyc and no nodes are shared between them except the end nodes of the shared link. The third condition is the value of redundancy[3] after aggregation of Cyc and $Cyc_i$ should be less than the redundancy before aggregation. After the aggregation we obtain one cycle that has one additional straddling link, and which can detours $n_i$ wavelengths along the cycle at the same time ($n_i$ take a value of link with a minimal working capacity $w_j$ ). The cycle obtained becomes the new Cyc that will be used in the next iteration. The process of aggregation continues until, one of the last conditions is violated. When the construction of efficient p-cycle is completed, the working capacities protected by this efficient p-cycle are removed.

The process of construction of p-cycles is repeated until all working capacity is protected or there are no more wavelengths in the network to construct p-cycle.

[3] The redundancy can be defined as total spare capacity divided by total working capacity

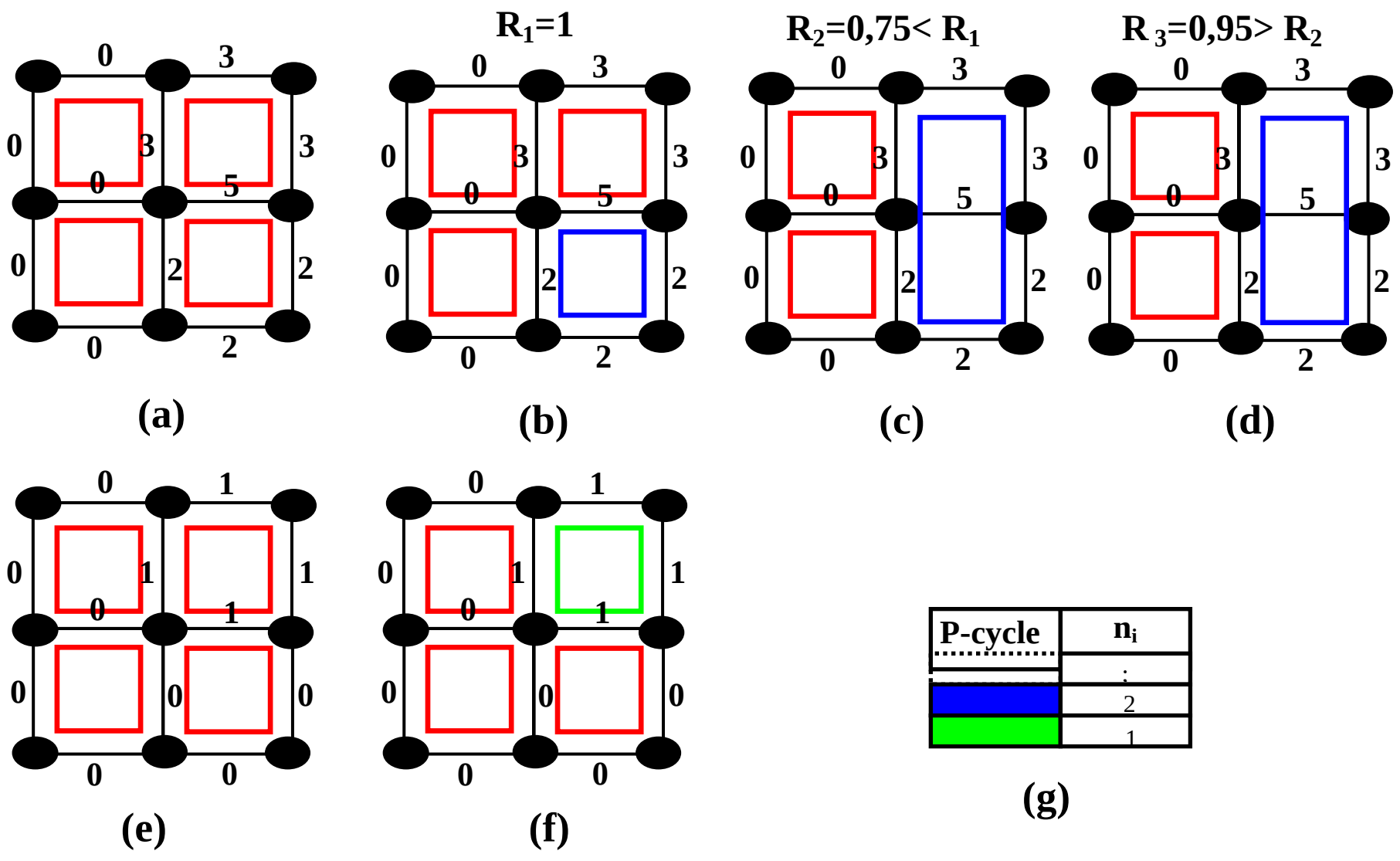

Fig. 5 High efficiency p-cycle construction process

Figure.5 (a) shows an optical network state after protecting some links. Figure.5 (b), (c), (d) shows the different phases of high efficiency p-cycle construction process. We can see in Fig. 5(d) that the aggregation is not possible because the third condition is not verified (the value of redundancy after aggregation should be less than the redundancy before aggregation). The p-cycle generated in phase (c) can detour 2 wavelengths along the cycle at the same time. The table shown in Fig.5 (g) contains some p-cycles which protect some links of network.

**Algorithm**

1. Select a cycle **Cyc** that has a link with a minimal $w_j$ ($w_j$>**0**) among all shortest cycles in the network. (If the link with a minimal $w_i$ is shared between two cycles. Then we choose the cycle that has more links unprotected).
2. Find a $cyc_i$ in the graph that satisfied the following conditions:
   a. ***Cyc*** and $Cyc_i$ share one and only one link between them.
   b. ***Cyc*** and $Cyc_i$ do not share nodes except the end nodes of the shared links.
   c. The value of ***redundancy*** after the aggregation of **Cyc** and $\mathbf{Cyc_i}$ is less than the redundancy before aggregation.
3. Perform an aggregation if a $Cyc_i$ is found. The new cycle becomes a new **Cyc**, then go to 2.
4. Remove the working capacity protected by **Cyc** (we remove $n_i$ wavelengths on any on-cycle link of the p-cycle and $2 \times n_i$ wavelengths on any straddling link of the p-cycle).
5. If there are working capacities unprotected then go to 2.
6. End.

## IV. Simulation Results and Analysis

In this section we evaluate our proposed algorithm in terms of redundancy and in terms of the size of final set of p-cycles that protect network. The network topology that we used to evaluate the performance of our solution is 28-node, 45-link USA long haul network taken from [9]. The working capacity in each link of network is obtained by routing the demand matrix over the shortest path. We assume each node in the network has full wavelength conversion capability. The test network is shown in Fig. 6.

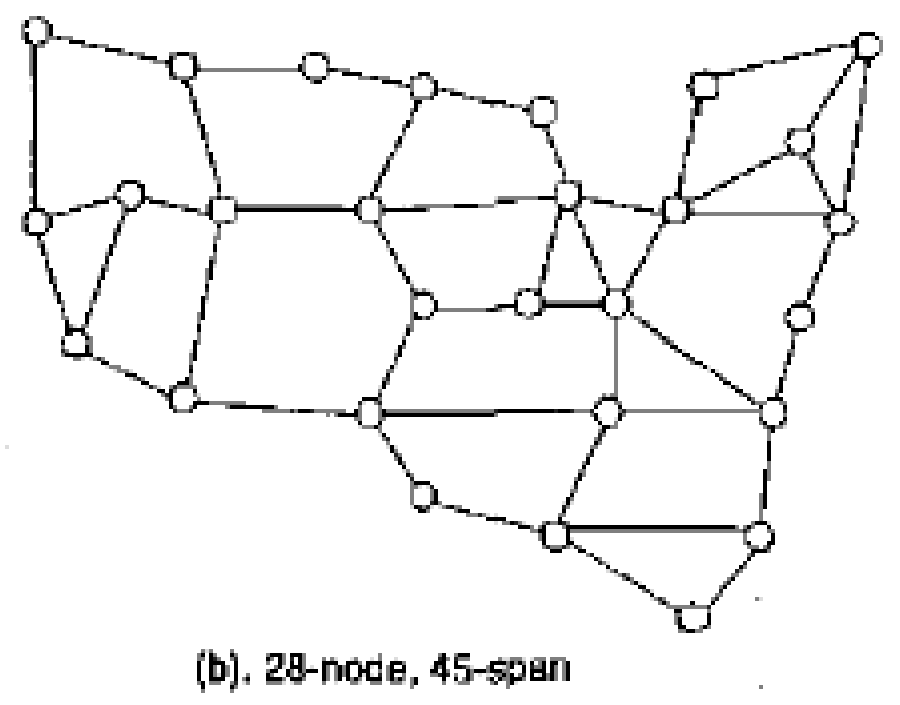

Fig. 6 Test network

We first test the performance of resources utilization (redundancy) for CIDA and for our algorithm in the above network. The redundancy can be defined as the total spare capacity divided by the total working capacity. Smaller value of redundancy means better performance of resource utilization.

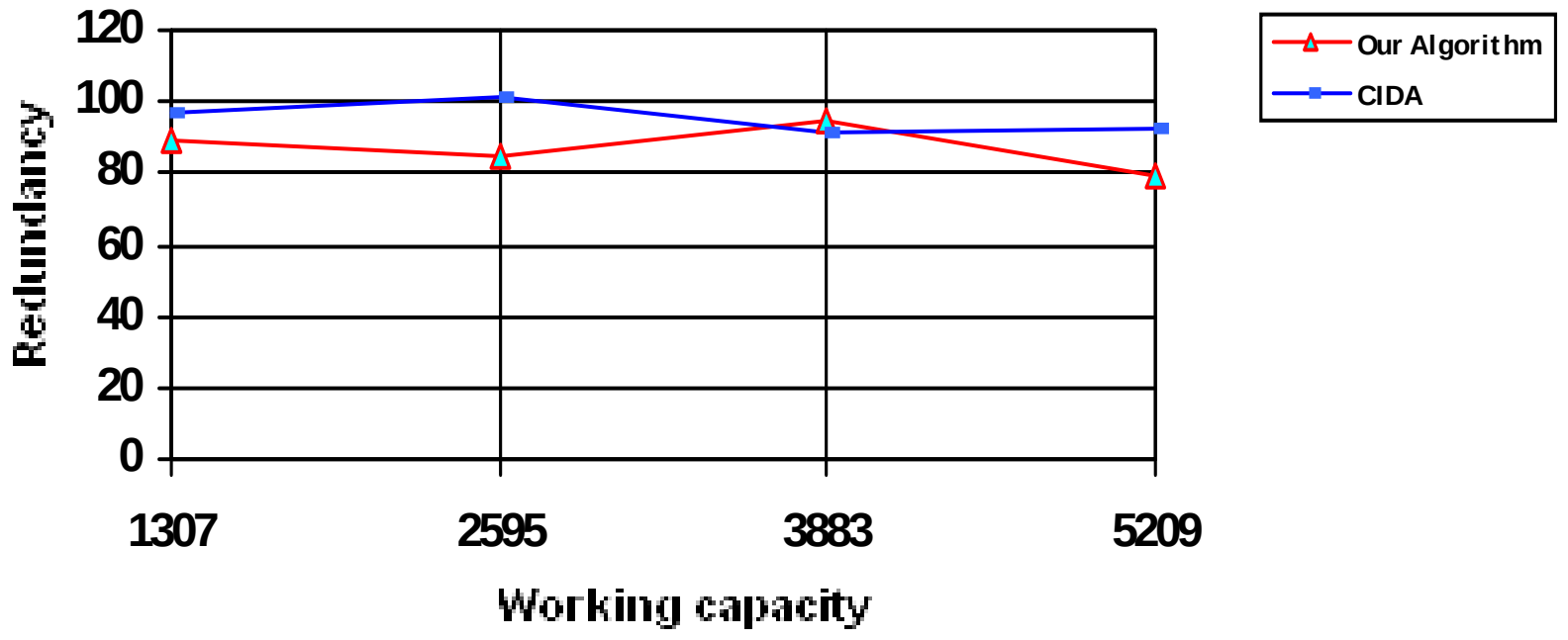

Fig. 7 Simulation comparing CIDA and our algorithm

We can see that the value of redundancy in our solution is always less than 100%. The reason for this is that our solution generates p-cycles by taking into account the state of the network. This enables us to construct only p-cycles that have small value of redundancy.

Since the number of p-cycles in the final set that protect network is an important performance for a configuration of WDM network. We also test the number of p-cycles used to protect the network. Smaller number of p-cycles in the final set of p-cycles means better performance

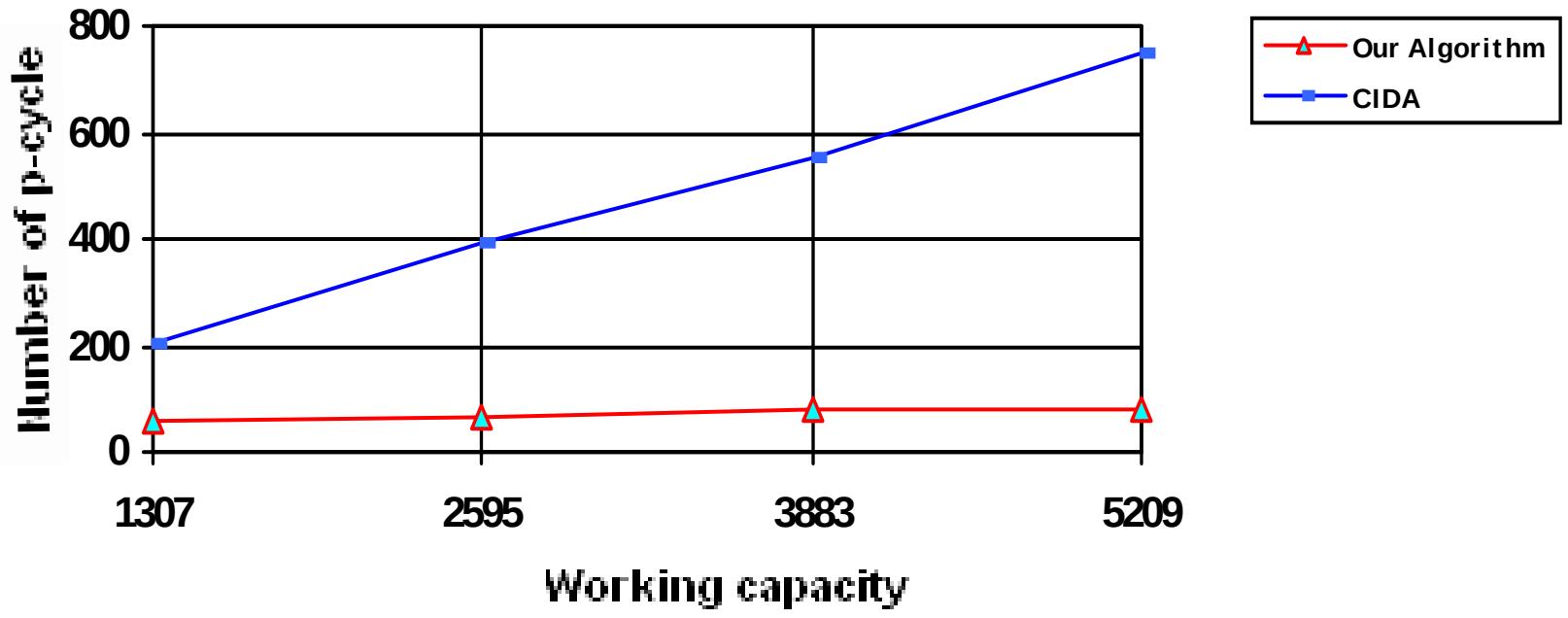

Fig. 8 Number of *p*-cycles selected to protect network

Simulation results of the number of p-cycles in the final set that protect network are summarized in Fig.8. We can see that our solution protects network with a small number of p-cycles. The reason for this is that in our algorithm the p-cycles generated can detour $n_i$ wavelengths along the cycle at the same time. But in CIDA all p-cycles detour one and only one wavelength along the cycle at the same time.

**V. Conclusion**

In this paper, we have addressed the protection in WDM optical networks using p-cycles. We have treated the problem of finding an optimal set of p-cycles that protect the optical network for a given working capacity. We have proposed a heuristic solution based an incremental aggregation of shortest cycle. Our generation of p-cycles takes into account the state of the network. This enables us to have a good value of redundancy and a very small number of p-cycles that protect network. The simulation has shown that the redundancy obtained by our heuristic algorithm is always less than 100%, and the number of p-cycles generated by our algorithm is very small than the number of p-cycles generated by CIDA.